\documentclass{INTERSPEECH2023}
\interspeechcameraready
\newcommand{\newpara}[1]{\vspace{4pt}\noindent\textbf{#1}}
\usepackage{tabularx,lipsum,xcolor}
\usepackage{amsmath}
\usepackage{multicol, multirow, booktabs, makecell}
\usepackage{cite}
\newcolumntype{Y}{>{\centering\arraybackslash}X}

\title{Encoder-decoder multimodal speaker change detection}
\name{Jee-weon Jung$^{1,2}$, Soonshin Seo$^{1,2}$, Hee-Soo Heo$^{2}$, Geonmin Kim$^2$,\\You Jin Kim$^{2}$, Young-ki Kwon$^{2}$, Minjae Lee$^2$, and Bong-Jin Lee$^{2}$}
\address{
  $^1$NAVER Corporation, Republic of Korea
  $^2$NAVER Cloud Corporation, Republic of Korea}
\email{jeeweon.jung@navercorp.com}

\begin{document}

\maketitle
 
\begin{abstract}
The task of speaker change detection (SCD), which detects points where speakers change in an input, is essential for several applications. Several studies solved the SCD task using audio inputs only and have shown limited performance. Recently, multimodal SCD (MMSCD) models, which utilise text modality in addition to audio, have shown improved performance. In this study, the proposed model are built upon two main proposals, a novel mechanism for modality fusion and the adoption of a encoder-decoder architecture. Different to previous MMSCD works that extract speaker embeddings from extremely short audio segments, aligned to a single word, we use a speaker embedding extracted from 1.5s. A transformer decoder layer further improves the performance of an encoder-only MMSCD model. The proposed model achieves state-of-the-art results among studies that report SCD performance and is also on par with recent work that combines SCD with automatic speech recognition via human transcription.
\end{abstract}

\section{Introduction}
\label{sec:intro}
Speaker change detection (SCD), also known as speaker segmentation, is a critical task that divides an input into multiple speaker-homogeneous segments by detecting the points where speakers change. SCD is essential in several applications, with speaker diarisation being the most common use case~\cite{hruz2017convolutional,bredin2020pyannote,cheng2009bic}. However, SCD can also be applied to various other tasks such as automatic speech recognition~\cite{liu1999fast}.

Pioneer works in the SCD literature have developed models using audio recordings as input~\cite{bredin2017speaker,malegaonkar2007efficient}. Various frameworks have been proposed to address this task. However, these systems have had limited performance. In our analysis, this is because audio-only SCD is a fundamentally difficult task to solve without additional information; we analyse audio-only SCD as a sequential speaker verification with extremely short segments. Despite recent advances in speaker recognition, stable performance can only be achieved for segments longer than one second. However, audio-only SCD needs to conduct speaker verification for every two consecutive frames where each frame typically conveys 10 to 20 ms of speech.

The challenge of audio-only SCD is further evident in current speaker diarisation solutions. Speaker diarisation, which solves ``who spoke when,'' comprises speaker segmentation (i.e., SCD) and clustering. Reliable SCD would allow us to apply clustering algorithms straightforwardly to SCD outputs. However, the majority of current solutions instead adopt endpoint detection and extract speaker embedding extractors from the voiced regions using a sliding window~\cite{park2022review,kwon2022multi}.

As a natural progression, researchers have investigated text-based or audio-text multimodal SCD (MMSCD) solutions~\cite{park2018multimodal}. Text-based and audio-text SCD models have shown to improve performance compared to audio-only models. However, reported ablations have shown that text contributes much more than the audio modality in the SCD task. Additionally, several studies only report diarisation performances, which hinders future researchers to compare and select which SCD to employ when constructing a process pipeline for different tasks using SCD.

In this study, we focus on MMSCD itself, excluding diarisation or other applications out of scope. We believe that if MMSCD becomes reliable and robust, it can be easily adopted for several other tasks, including speaker diarisation. Compared to other multimodal combinations, audio-text is relatively easy to implement due to advances in automatic speech recognition that transcribe and derive text from speech\footnote{It is worth noting that this paper deals with ground truth transcriptions. In real-world scenarios, these transcriptions should be replaced by automatic speech recognition systems' outputs.}.

Our proposed MMSCD model is built upon two main proposals. Firstly, we analyse the limitations of existing modality fusion mechanisms and propose a novel mechanism that enables audio modality to contribute more. Secondly, we propose to adopt the encoder-decoder (i.e., sequence-to-sequence (seq2seq)) architecture instead of encoder-only architecture in SCD for the first time and prove its effectiveness. We conduct experiments using two widely adopted datasets in the literature: Switchboard1~\cite{godfrey1993switchboard} and AMI corpus~\cite{carletta2007unleashing}. Our proposed model demonstrates competitive performance, achieving state-of-the-art results among studies that report SCD performance. Our model is also on par with another line of recent work in SCD that combines SCD with automatic speech recognition via large-scale human transcription~\cite{zhao2022augmenting}.

The rest of this paper is organised as follows. Section~\ref{sec:proposed} describes the proposed MMSCD model. Experiments and corresponding results are discussed in Sections \ref{sec:exp} and \ref{sec:results}, followed by a conclusion.

\section{Proposed MMSCD model}
\label{sec:proposed}
\begin{figure*}[t!]
    \centering
    \includegraphics[width=0.8\linewidth]{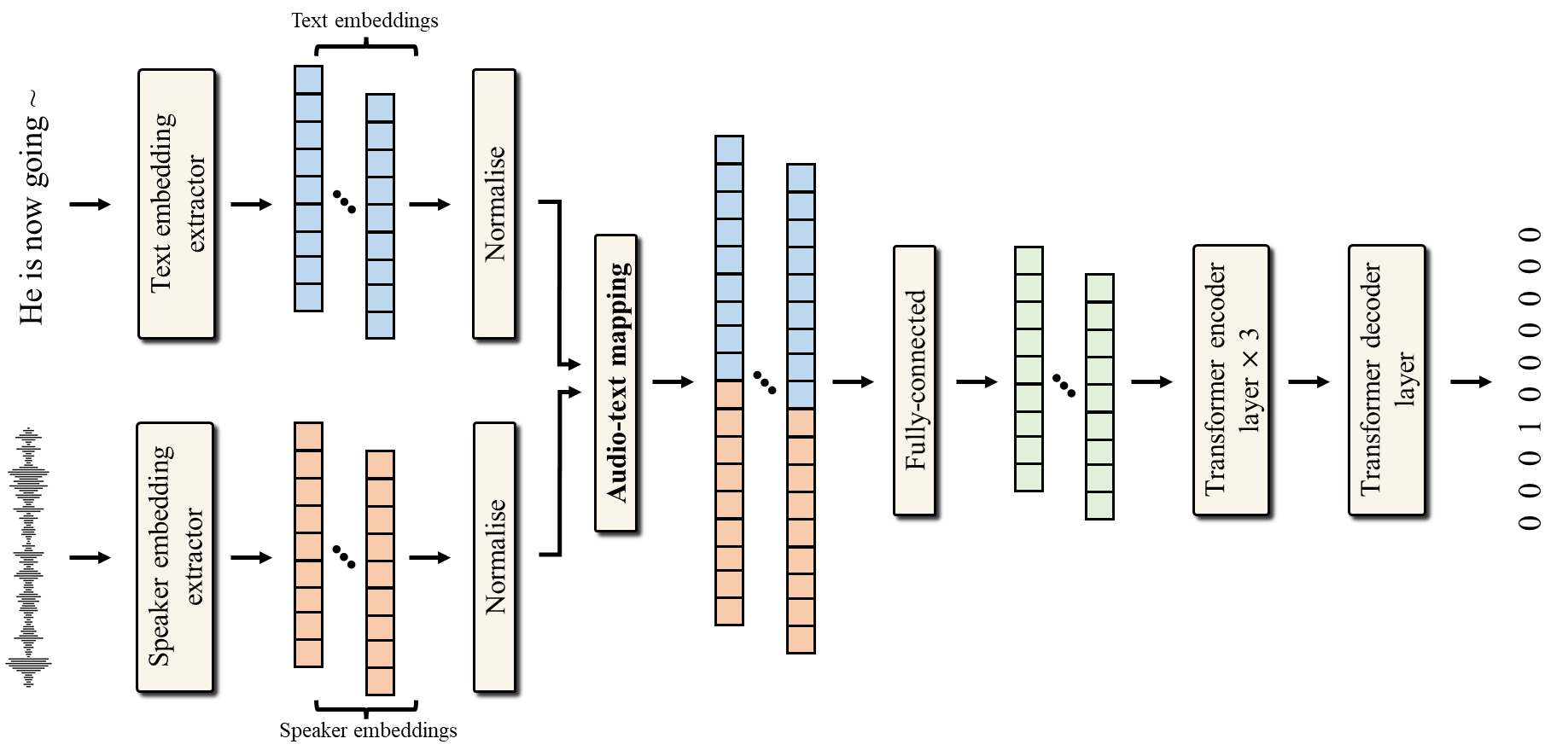}
    \caption{Overall framework of the proposed multimodal speaker change detection (MMSCD).}
    \label{fig:framework}
    \vspace{-5pt}
\end{figure*}

\begin{figure}[t!]
    \centering
    \includegraphics[width=\columnwidth,height=5cm]{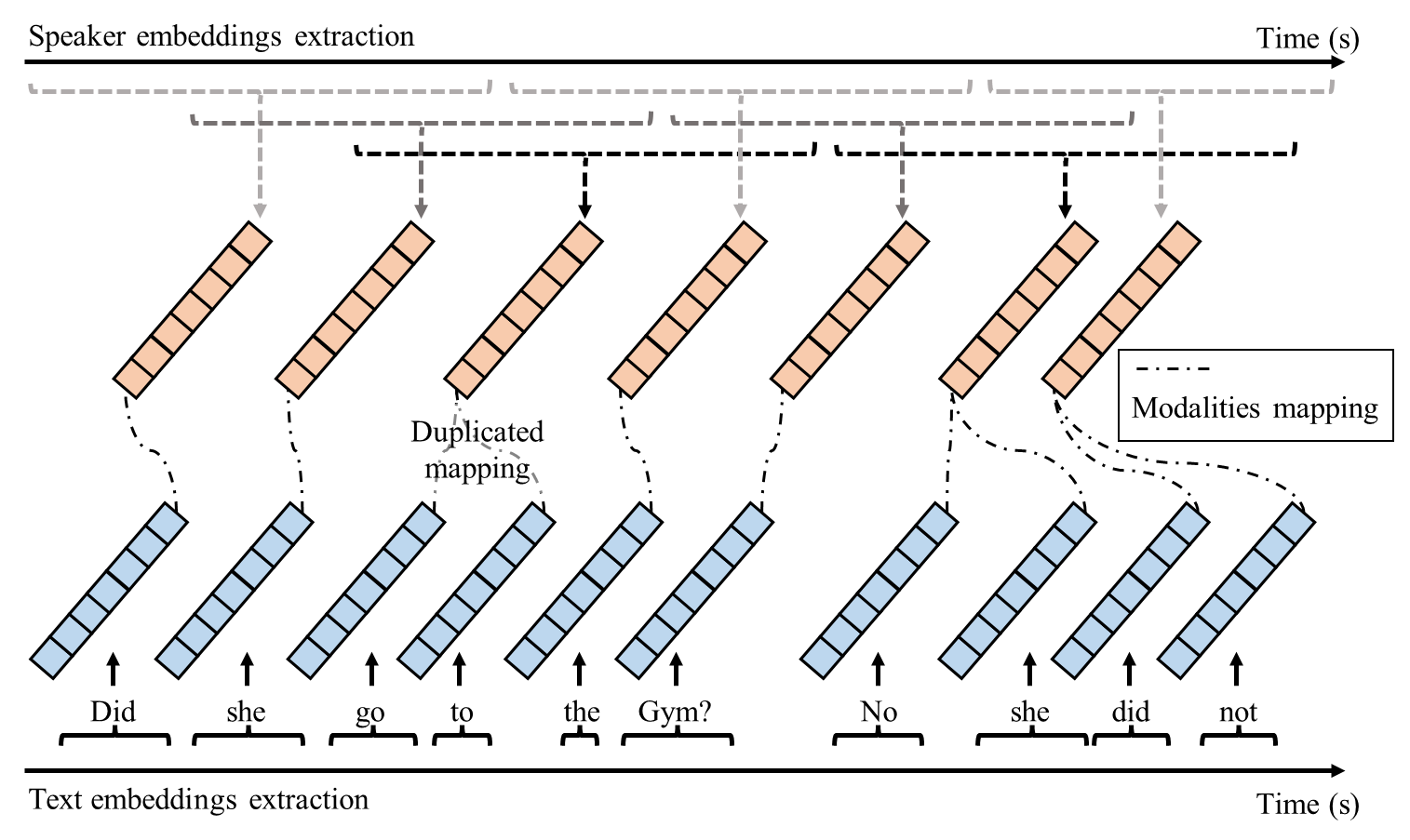}
    \caption{Proposed audio-text mapping for MMSCD input. Different from existing methods that extract speaker embedding using the onset and offset of a word, we extract speaker embeddings in a sliding window fashion and then map a speaker embedding closest to the corresponding word.}
    \label{fig:atMapping}
    \vspace{-5pt}
\end{figure}

The overall framework of our MMSCD model is illustrated in Figure~\ref{fig:framework}.

\subsection{Prerequisites: single-modal embedding extractors}

Our proposed MMSCD model is built upon two single-modal embedding extractors each of which extracts speaker and text embeddings from an input. 

\newpara{Speaker embedding extractor.}
We utilise the speaker embedding extractor proposed in \cite{jung2023search,jung2022large}. It adopts the MFA-Conformer~\cite{zhang2022mfa} architecture, a variant of the Conformer~\cite{gulati2020conformer} model. MFA-Conformer does not modify the Conformer encoders, but concatenates all encoder outputs and then employs an attentive statistics pooling~\cite{okabe2018attentive} sequentially.
The model is trained using a massive amount of labelled data and with data augmentations designed to handle multiple speakers being present in a short audio segment. The model extracts 256-dimensional speaker embedding from a segment of 1.5s with a shift size of 0.5s.

\newpara{RoBERTa.}
We adopt RoBERTa, proposed in \cite{liu2019roberta}, as the text embedding extractor. We use the pre-trained RoBERTa-base model\footnote{Available at: \url{https://huggingface.co/roberta-base}.}, without further modification or fine-tuning. The model extracts 768-dimensional text embeddings from each sub-word input. 

\subsection{Modality fusion with stable speaker embeddings}
\label{ssec:modalfusion}
Integrating two different modalities is one of the most critical components of multimodal models. Several strategies have been explored~\cite{nagrani2021attention}. In terms of timing, both early and late fusions have been reported to be effective depending on different cases, with early-fusion being more dominant. In the case of MMSCD, previous works~\cite{park2018multimodal,park2019speaker} have mostly adopted early-fusion. We also choose early fusion following the majority of preceding works.

We focus on the unit of each modality when composing an input for the MMSCD model. Previous works have adopted the word token as a unit, as it is a competitive candidate, compared to ``{\em audio frames},'' where the typical shift size is 10ms or 20ms. By using word (or sub-word) as the input unit, audio or speaker embeddings can be extracted from extremely short durations; the duration of a typical word is no longer than 300ms. However, speaker embeddings extracted from such a short duration are highly vulnerable; typically a duration of 3 seconds or longer is required in speaker recognition. Although speaker embedding extraction has experienced rapid breakthroughs, extracting embeddings from a segment less than 2 or 1.5 seconds are is a well known cause of performance degradation ~\cite{tawara2020frame, jung2019short}.

We propose a novel audio-text mapping mechanism for the MMSCD task, illustrated in Figure~\ref{fig:atMapping}, inspired by the speech segment selection process of \cite{park2019speaker}.
Firstly, speaker embeddings are derived in a sliding window fashion with a 1.5s window and 0.5s shift. When composing an audio-text pair input for the MMSCD model, the speaker embedding which has the closest midpoint to the midpoint of the text in terms of time is selected. Since the proposed mechanism extracts speaker embedding from a longer duration, the information from audio modality can be more reliable and stable. There is a limitation that a few (typically, two) tokens may have identical speaker embedding as a pair. Nevertheless, this should not negatively affect the performance since our system makes the decision token-wise, meaning that our framework assumes audio modality complementing text modality.

The specific process for fusing the two embeddings before feeding them to encoder layers is as follows. We first extract speaker and text embeddings. Each embedding is normalised to have a magnitude of $sqrt(dim)$ following Kaldi recipe~\cite{Povey_ASRU2011}, where $dim$ is the dimensionality, and then concatenated. Next, a fully-connected layer projects the concatenated vector into another space, thereby fusing the two embeddings. Finally, we add positional encoding to the fused embedding.

\subsection{Seq2seq model architecture}
\label{ssec:architecture}
Various deep neural network (DNN) architectures have been explored for the SCD task. In studies of both audio-only and multimodal SCD~\cite{wang2017speaker,bredin2021end, anidjar2021hybrid, sari2019pre}, fully-connected, long short-term memory, or gated recurrent unit layers have been widely utilised. Transducer architectures~\cite{graves2012sequence, zhang2020transformer}, another widely utilised automatic speech recognition framework, have been recently explored in the SCD task.
Yet, compared to automatic speech recognition or other seq2seq tasks, existing works only have leveraged encoder layers, remaining decoder layers unexplored. 

We aim to investigate whether employing decoder layers to the SCD encoder output can further improve the performance of the MMSCD model.
Our underlying assumption is that posing inductive bias to consider past frames and predicting whether the speaker has changed in the current timestep (i.e., autoregressive decoding) may be advantageous for SCD, as it has been proven effective in automatic speech recognition and other tasks.

To extend the model architecture from encoder-only to encoder-decoder~\cite{chorowski2015attention,chan2016listen}, we make the following modifications. Firstly, we add a \texttt{<bos>} token before the first token. For the corresponding speaker embedding, we duplicate the first speaker embedding of the input. For the label, we append \texttt{<eos>} token at the end. Note that during the inference phase, we disregard the \texttt{<eos>} token and infer a pre-defined length.
Furthermore, we explore whether beam search decoding~\cite{tu2017neural,freitag2017beam} can enhance the performance of encoder-decoder models.

\newpara{Autoregressive training.}
In seq2seq models, a discrepancy exists between training and inference phases. During the training phase, teacher forcing~\cite{ronald1989learning} is commonly used as it helps to increase the convergence speed and performance. In this approach, the ground truth labels are fed as input at each timestep instead of model prediction from the previous timestep. However, in the inference phase, such mechanism is not possible, and the model relies upon its own prediction from the previous step.
Since the model has no experience of having made incorrect predictions in the previous timesteps, error accumulation caused by recurrent prediction may significantly deteriorate~\cite{liu2019sequence, zeng2022transformers}. This effect may be more pronounced in the SCD task since the output is binary and imbalanced.
Drawing inspiration from \cite{liang2022janus,ghazvininejad2020semi}, we therefore investigate the possibility of training the model in an autoregressive manner from a certain epoch and compare the results.

\section{Experiments}
\label{sec:exp}
\begin{table}[t]
    \caption{Statistics of train and evaluation sets of used datasets.}
    \label{tab:dbStat}
    \centering
    \begin{tabularx}{\linewidth}{lYYY}
      \toprule
      Dataset & Domain & Trn dur (h) & Eval dur (h)\\
      \toprule
      Switchboard1 & Telephone & 248.45 & 10.68\\
      AMI & Meeting & 80.67 & 9.06\\
      \bottomrule
    \end{tabularx}
\end{table}

\subsection{Datasets and metrics}
\label{ssec:dbMetric}
Two widely adopted datasets are used for verifying the proposals of this study: Switchboard1-release 2~\cite{godfrey1993switchboard} and AMI corpus Mixed headset~\cite{carletta2007unleashing}. For the Switchboard1 dataset, we upsample all recordings to 16kHz and mix them into mono recordings. In addition, since Switchboard1 does not have an official train and evaluation split, we randomly select 100 audio recordings for evaluation~\footnote{Certain number of seen speakers may exist in the evaluation set because Switchboard1 dataset has reoccurring speakers.}. Table~\ref{tab:dbStat} describes the duration specifications of the two datasets. We adopt F1 score and equal error rate (EER) as the primary and the supplementary metrics. EER is used because label imbalance is severe and it equally weights false alarm and false rejections~\footnote{For both datasets, ``{\em speaker change}'' label occupies less than 10\%.}. 

\subsection{Configurations}

\newpara{Speaker embedding extractor.}
We train the MFA-Conformer model using the {\em \textbf{v4}} configuration of \cite{jung2022large}, which includes 88k speakers' 10k hours of speech.
It adopts six 512-dimensional Conformer encoder blocks with eight attention heads.
80-dimensional log mel-spectrograms extracted every 10ms using a 25ms window are fed to the model.
During the training phase, data augmentations from \cite{jung2023search} are adopted as well as adding noise and reverberations.

\newpara{MMSCD model.}
The encoder input projection module, which digests concatenated speaker and text embeddings, consists of a fully-connected layer with 512 dimensions, followed by dropout~\cite{srivastava2014dropout} and a GELU~\cite{hendrycks2016gaussian} non-linearity.
The encoder part includes three Transformer~\cite{vaswani2017attention} layers with 512 dimensions and eight multi-heads for attention.
The decoder input projection module comprises identical layers to the encoder input projection module. 
The only difference is the input dimension; encoder projection module inputs 1024-dimensional vector (256 for speaker and 768 for text) whereas decoder projection module inputs 768-dimensional vector because it only digests text embedding indicating same of different speakers.
The decoder part comprises one Transformer decoder layer with 512 dimensions and eight attention heads.

\newpara{Training details.}
Models trained for either of the two datasets employ AdamW~\cite{ilya2019decoupled} optimiser with an initial learning rate of $1e^{-3}$ and a weight decay of $5e^{-5}$.
We adopt a cosine learning rate scheduler with a minimum learning rate of $5e^{-6}$. Learning rate warm-up is also adopted during the first 1,000 iterations. For models trained on the AMI corpus, the batch size is 32 and the number of epochs is 8,000~\footnote{This large number of epochs is due to the limited number of samples of the AMI train set, which is 136.}. Models trained on the Switchboard corpus adopt a batch size of 64 and 400 epochs.

\begin{table*}[t]
    \caption{Main results. Through comparison experiments conducted on Switchboard1 and AMI datasets, it is shown that adopting seq2seq architecture with a decoder layer, instead of composing an encoder-only architecture is beneficial. We also demonstrate that both modalities contribute to performance, where the audio modality also plays a crucial role.}
    \label{tab:mainRes}
    \centering
    \begin{tabularx}{\linewidth}{l|YYYY|YYYY}
      \toprule
      & \multicolumn{4}{c|}{\textbf{Switchboard1}} & \multicolumn{4}{c}{\textbf{AMI}}\\
      \cmidrule{2-9}
      & \textbf{Prec} & \textbf{Recall} & \textbf{F1} & \textbf{EER} & \textbf{Prec} & \textbf{Recall} & \textbf{F1} & \textbf{EER}\\
      \toprule
      \textbf{\textit{Proposed}} & 82.28 & 83.09 & 82.68 & 5.90 & 66.73 & 80.49 & 73.61 & 13.00\\
      \hspace{3mm}w/o decoder & 80.84 & 80.62 & 80.73 & 6.36 & 57.33 & 78.76 & 68.05 & 15.97\\
      \hspace{3mm}w/o audio modality & 80.59 & 68.69 & 74.60 & 10.45 &  61.51 & 76.72 & 69.12 & 15.13\\
      \hspace{3mm}w/o text modality & 25.65 & 88.56 & 57.10 & 17.53 & 20.01 & 96.77 & 58.39 & 27.34\\
      \bottomrule
    \end{tabularx}
    \vspace{-5pt}
\end{table*}
\begin{table}[t]
    \caption{Ablations on the effect of autoregressive training and its starting point. Evaluated on Switchboard1. Total number of epochs is 400 (TF: teacher forcing, AR: autoregressive training).}
    \label{tab:autoreg}
    \centering
    \begin{tabularx}{\linewidth}{lYYYY}
      \toprule
      \textbf{\# TF/AR epochs} &  \textbf{Prec} & \textbf{Recall} & \textbf{F1} & \textbf{EER}\\
      \toprule
      400/0 & 77.94 & 80.63 & 79.28 & 6.63\\
      200/200 & \textbf{82.91} & 82.84 & \textbf{82.87} & 5.94\\
      300/100 & 82.28 & \textbf{83.09} & 82.68 & \textbf{5.90}\\
      350/50 & 83.79 & 81.35 & 82.57 & 5.94\\
      \bottomrule
    \end{tabularx}
    \vspace{-15pt}
\end{table}
\begin{table}[t]
    \caption{Exploration of beam search. Results reported on Switchboard1. Performance of greedy equals to beam search with width=1 (w: beam width).}
    \label{tab:beamsearch}
    \centering
    \begin{tabularx}{\linewidth}{lcYYYY}
      \toprule
       & decoding & Prec & Recall & F1 & EER\\
      \toprule
      \multirow{2}{*}{w/o autoreg trn} & greedy & 77.94 & 80.63 & 79.28 & 6.63\\
      & beam(w=100) & \textbf{78.87} & \textbf{81.72} & \textbf{80.30} & \textbf{6.44}\\
      \midrule
      \multirow{2}{*}{w autoreg trn}& greedy & \textbf{82.28} & 83.09 & 82.68 & \textbf{5.90}\\
      & beam(w=100) & 82.26 & \textbf{83.14} & \textbf{82.70} & 5.91\\
      \bottomrule
    \end{tabularx}
    \vspace{-5pt}
\end{table}
\begin{table}[t]
    \caption{Comparison with other works. Direct comparisons should be avoided. Refer to explanation in the main text for full details (NR: not reported).}
    \label{tab:otherWorks}
    \centering
    \begin{tabularx}{\linewidth}{lYYYY}
      \toprule
       &  \textbf{Prec} & \textbf{Recall} & \textbf{F1} & \textbf{EER}\\
      \toprule
      \multicolumn{5}{c}{\textbf{Switchboard1}}\\
      \toprule
      India et al.~\cite{india2017lstm} & 62.54 & 66.70 & 64.55 & NR\\
      \textbf{\textit{Ours}} & 82.28 & 83.09 & 82.68 & 5.90\\
      \toprule
      \multicolumn{5}{c}{\textbf{AMI}}\\
      \toprule
      Zhao et al.~\cite{zhao2022augmenting} & 79.40 & 68.1 & 73.30 & NR\\
      \textbf{\textit{Ours}} & 66.73 & 80.49 & 73.61 & 13.00\\
      \bottomrule
    \end{tabularx}
    \vspace{-15pt}
\end{table}
\section{Results}
\label{sec:results}
\subsection{Main results}
\label{ssec:mainres}
Table~\ref{tab:mainRes} addresses the main results on two datasets. Firstly, we confirm that including a decoder layer for the SCD task can be advantageous. In both datasets, the inclusion of a decoder layer improved both the F1 score and EER. For instance, in AMI, the F1 score increased from 68.05 to 73.61. 

Next, we observe that training and inferring without either modality leads to severe performance degradation. Removing the audio modality significantly deteriorates the performance in our model, with EER on Switchboard1 increasing from 5.90\% to 10.45\%. This finding contrasts with previous works which report that the audio modality only has a minor contribution to the performance of MMSCD~\cite{park2019speaker}. Therefore, we conclude that the proposed modality matching mechanism is effective.

\subsection{Autoregressive training}
Table~\ref{tab:autoreg} describes experiments on the effect of autoregressive training. The first row addresses performance without autoregressive training. The other rows depict the result when the training starts with teacher forcing but trains the model in an autoregressive manner during a certain period. The results show that regardless of the starting timing, autoregressive training improves the performance by more than 5\%. 
However, the performance difference between different starting epochs was negligible. Hence it may be a good strategy to begin autoregressive training at the last moment, since autoregressive training slows down the training phase.

\subsection{Exploration on beam search}
\label{ssec:beamsearch}
In automatic speech recognition, applying the beam search algorithm~\cite{graves2014towards,hannun2014deep}, often combined with language models, can significantly improve the performance. We apply a typical beam search; however, we leave all nodes for the first few timesteps, since the SCD task has only two labels (i.e., speaker change and non-change), and thus the total beam width can be smaller than the allowed beam width.
Table~\ref{tab:beamsearch} describes the result of our exploration of beam search decoding. When autoregressive training is not applied, beam search was effective in decreasing the EER and increasing the F1 score. However, when autoregressive training was applied, beam search could not bring further improvements. We believe that beam search has future potentials worth exploring; however, at this stage, it only contributes when the performance is less stable.

\subsection{Comparison with other works}
\label{ssec:comparison_literature}
Table~\ref{tab:otherWorks} compares our work with previous works in the literature.
However, direct comparisons are not available due to a few different configurations.
For comparison on Switchboard1, the authors of \cite{india2017lstm} composed an evaluation set with mixed Fisher and Switchboard1 datasets. They also adopt forgiveness collars of 500ms and the unit is audio frames.
For comparison on AMI, the authors of \cite{zhao2022augmenting} adopt a custom definition of both precision and recall which makes the task easier. They also adopt forgiveness collars of 250ms and the unit is audio frames.
In contrast, our unit is sub-word and we do not adopt any forgiveness collars, hence we argue that our evaluation is the most stringent.
Although direct comparisons are not available, even with much more stringent evaluation setting, our model outperforms \cite{india2017lstm} by a large margin in Switchboard1 and demonstrates on par performance with \cite{zhao2022augmenting} in AMI.

\section{Conclusion}
\label{sec:conclusion}
We conducted an investigation into the MMSCD task, validating two main hypotheses. Firstly, we assumed that incorporating decoder layers into an encoder-decoder architecture would be advantageous for the SCD task. We demonstrated that including decoder layers can significantly enhance performance by almost 10\%. Secondly, we proposed a novel mechanism for using speaker embeddings. Rather than extracting speaker embeddings using the onset and offset of words, we extracted them using a sliding window and then mapped them to the corresponding words. This mechanism produced more stable speaker embeddings, as they have been extracted from  longer durations. Through experiments on two widely-used datasets, we confirmed competitive performance, outperforming previous audio-only and MMSCD models and achieving comparable performance with the recent automatic speech recognition re-annotation-based SCD method.

\clearpage
\bibliographystyle{IEEEtran}
\bibliography{shortstrings,refs}

\end{document}